\documentclass[twocolumn]{jpsj2} 
\usepackage{bm}
\usepackage{color}

\title{High-field Phase Diagram and Spin Structure of Volborthite Cu$_3$V$_2$O$_7$(OH)$_2 \cdot $2H$_2$O}

\author{Makoto \textsc{Yoshida}\thanks{E-mail address: yopida@issp.u-tokyo.ac.jp}, 
Masashi \textsc{Takigawa}\thanks{E-mail address: masashi@issp.u-tokyo.ac.jp}, 
Steffen \textsc{Kr\"{a}mer}$^{1}$, 
Sutirtha \textsc{Mukhopadhyay}$^{1}$, 
Mladen \textsc{Horvati\'{c}}$^{1}$, 
Claude \textsc{Berthier}$^{1}$, 
Hiroyuki \textsc{Yoshida}$^{2}$, 
Yoshihiko \textsc{Okamoto}, 
and Zenji \textsc{Hiroi}}

\inst{Institute for Solid State Physics, University of Tokyo, Kashiwa, Chiba 277-8581, Japan \\ 
$^{1}$Laboratoire National des Champs Magnetique Intenses, LNCMI - CNRS (UPR3228), UJF, UPS and INSA, BP 166, 38042 Grenoble Cedex 9, France\\
$^{2}$National Institute for Materials Science (NIMS), 1-1 Namiki, Tsukuba, Ibaraki 305-0044, Japan}

\abst{We report results of $^{51}$V NMR experiments on a high-quality powder 
sample of volborthite Cu$_3$V$_2$O$_7$(OH)$_2 \cdot $2H$_2$O, 
a spin-1/2 Heisenberg antiferromagnet on a distorted kagome lattice. 
Following the previous experiments in magnetic fields $B$ below 12 T, 
the NMR measurements have been extended to higher fields up to 31 T. 
In addition to the two already known ordered phases (phases I and II), 
we found a new high-field phase (phase III) above 25~T, at which a second 
magnetization step has been observed. 
The transition from the paramagnetic phase to the antiferromagnetic phase III 
occurs at 26 K, which is much higher than the 
transition temperatures from the paramagnetic to the lower field phases I ($B$ $<$ 4.5~T) 
and II (4.5 $<$ $B$ $<$ 25~T). At low temperatures, two types of the V sites are 
observed with different relaxation rates and line shapes in phase III 
as well as in phase II. Our results indicate that both phases II and III 
exhibit a heterogeneous spin state consisting of two spatially alternating 
Cu spin systems, one of which exhibits anomalous spin fluctuations contrasting 
with the other showing a conventional static order. 
The magnetization of the latter system exhibits a sudden increase upon 
entering into phase III, resulting in the second magnetization step at 26 T.
We discuss the possible spin structure in phase III.}

\kword{kagome lattice, frustration, volborthite, Cu$_3$V$_2$O$_7$(OH)$_2 \cdot $2H$_2$O, NMR, spin structure, phase diagram, high-field}

\begin{document}
\maketitle

\section{Introduction} 
\label{introduction}
The kagome lattice, a two-dimensional (2D) network of corner-sharing equilateral triangles, 
is known for strong geometrical frustration. In particular, the ground state of the quantum 
$S$ = 1/2 Heisenberg model with a nearest-neighbor antiferromagnetic (AF) interaction on the 
kagome lattice is believed to show no long-range magnetic order. Theories have proposed various 
ground states such as spin liquids with no broken symmetry with \cite{Waldtmann} or without 
\cite{Hermele} a spin-gap or symmetry breaking valence-bond-crystal states \cite{Singh}. 
Real materials, however, have secondary interactions, which may stabilize a magnetic order. 
Effects of the Dzyaloshinsky-Moriya (DM) interaction \cite{Cepas}, spatially anisotropic 
exchange interactions \cite{Schnyder,Wang,Stoudenmire}, and longer range Heisenberg interactions \cite{Domenge} 
have been theoretically investigated. In the classical system, 
it is known that order-by-disorder effect \cite{Reimers} or DM interaction \cite{Elhajal} favors 
long-range order with the $\sqrt{3} \times  \sqrt{3}$ pattern or 
the ${\bf Q}$ = 0 propagation vector, respectively. 

\begin{figure}[b]
\begin{center}
\includegraphics[width=0.7\linewidth]{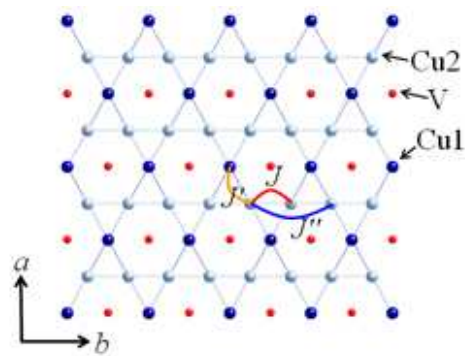}
\end{center}
\caption{(Color online) Schematic structure of volborthite projected onto the $a$-$b$ plane. 
The H and O sites are not shown. The V sites are located both below and above the Cu kagome layers. 
The upper and lower V sites are related by inversion with respect to the Cu sites, hence all V sites 
are equivalent.}
\label{fig1}
\end{figure}

On the experimental side, intensive efforts have been devoted to 
synthesize materials containing $S$ = 1/2 spins on a kagome lattice \cite{Hiroi,Shores,Okamoto,Morita}. 
Candidate materials known to date, however, depart from the ideal kagome model 
in one way or another, such as disorder, structural distortion, anisotropy, 
or longer range interactions. For example, herbertsmithite ZnCu$_3$(OH)$_6$Cl$_2$ 
realizes a structurally ideal kagome lattice \cite{Shores} and exhibits no magnetic 
order down to 50~mK \cite{Helton}. However, chemical disorder replacing 
10\% of Cu$^{2+}$ spins with nonmagnetic Zn$^{2+}$ ions \cite{Lee} significantly disturbs 
the intrinsic properties at low temperatures \cite{Imai,Olariu}. 

Volborthite Cu$_3$V$_2$O$_7$(OH)$_2 \cdot $2H$_2$O has distorted kagome layers formed 
by isosceles triangles. Consequently, it has two inequivalent Cu sites and two kinds of exchange 
interactions ($J$ and $J'$ ) as shown in Fig.~\ref{fig1}. The Cu2 sites form linear chains, 
which are connected through the Cu1 sites. The magnetic susceptibility $\chi $ obeys the 
Curie-Weiss law $\chi  = C/(T + \theta _W)$ above 200~K with $\theta _W$ = 115~K, 
exhibits a broad maximum at 20~K, and approaches a finite value 
at the lowest temperatures \cite{Hiroi,HYoshida}. An unusual magnetic transition 
has been observed near 1~K \cite{Fukaya,Bert,MYoshida1,Yamashita}, which is much lower than $\theta _W$, 
consistent with strong frustration in a kagome lattice. However, a recent density functional 
calculation \cite{Janson} proposed that the frustration is attributed to the competition 
between a ferromagnetic $J$ and an antiferromagnetic 
$J''$ between second neighbors along the chain (Fig.~1) 
rather than the geometry of a kagome lattice. 
Thus the appropriate spin model for volborthite has not been settled yet. 
Recently, anomalous sequential magnetization steps were reported in a 
high quality polycrystalline sample at 4.3, 25.5, and 46~T \cite{HYoshida}. 
Furthermore, a magnetization plateau was observed at 2/5 of the saturation magnetization
above 60~T \cite{Okamoto2}, in spite of the theoretical prediction for 
a plateau \cite{Hida} or a ramp \cite{Nakano} at the 1/3 of the saturation magnetization
in an isotropic kagome lattice. 

The magnetic transition at $T^* \sim 1$~K has been detected by $^{51}$V NMR \cite{Bert,MYoshida1}, 
muon spin relaxation \cite{Fukaya}, and heat capacity \cite{Yamashita} experiments. 
The NMR measurements revealed a sharp peak in the nuclear relaxation rate 1/$T_1$ 
and broadening of the NMR spectrum due to development of spontaneous moments for 
the magnetic field $B$ below 4.5~T \cite{MYoshida1}. A kink was also observed in the heat capacity \cite{Yamashita}. 
Recently, development of short range spin correlation was detected by neutron inelastic 
scattering experiments \cite{Nilsen}. However, the low $T$ phase, which we call phase I, 
shows various anomalies incompatible with a conventional magnetic order \cite{MYoshida1}. 
The NMR line shape is not rectangular but can be fit to a Lorentzian, 
suggesting a state in which the amplitude of the static moment has spatial modulation, 
such as a spin density wave (SDW) state or a spatially disordered state. 
The behavior 1/$T_1 \propto  T$ provides evidence for dense low-energy excitations. 
The spin-echo decay rate 1/$T_2$ is anomalously large, pointing to unusually slow fluctuations. 

Above 4.5~T, at which the first magnetization step occurs, 
another magnetic phase appears with a different spin structure and dynamics \cite{MYoshida1,MYoshida2}. 
In this phase, which we call phase II, the NMR results show coexistence of two types of the V sites 
with different values of 1/$T_2$ and line shapes \cite{MYoshida2}. 
The sites with large 1/$T_2$ have similar features 
to the V sites in phase I. They are strongly coupled to the Cu spins with a spatially modulated structure
and large temporal fluctuations. On the other hand, the sites with small 1/$T_2$ show a rectangular 
spectral shape, which is compatible with a conventional order having a fixed magnitude of the ordered moment, 
such as a N\'{e}el state or a spiral order. 
The numbers of these two sites are nearly equal 
in a wide range of $B$ and $T$. Based on these observations,
a heterogeneous spin state, in which two types of Cu moments form a periodic structure,
has been proposed \cite{MYoshida2}. 

\begin{figure}[tb]
\begin{center}
\includegraphics[width=0.9\linewidth]{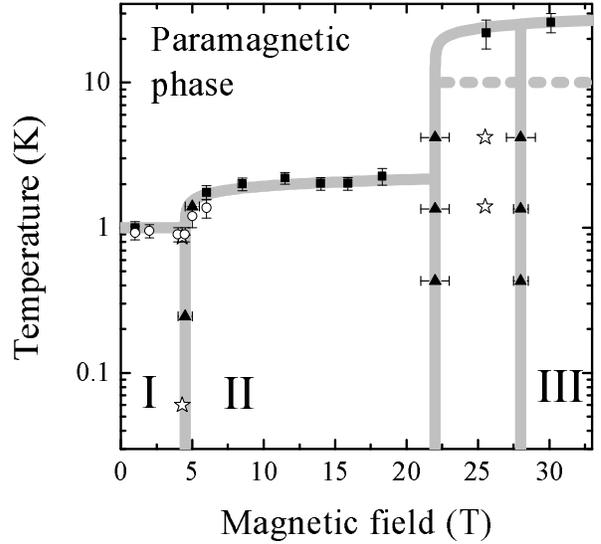}
\end{center}
\caption{The phase diagram of volborthite proposed from the previous \cite{HYoshida,MYoshida1} 
and present studies. The squares (triangles) represent the phase boundaries determined from the 
variation of the NMR spectra as a function of temperature (magnetic field). 
The circles and stars indicate the peak of 1/$T_1$ \cite{MYoshida1} and 
magnetization steps \cite{HYoshida}, respectively. The lines are guide to the eyes.
There is a finite range of field, where phase II and III coexist. 
}
\label{fig2}
\end{figure}

In this paper, we report results of the NMR experiments in higher magnetic fields up to 31 T. 
The phase diagram determined from the previous \cite{MYoshida1,MYoshida2} and the present 
experiments is shown in Fig.~\ref{fig2}.
A new high-field phase defined as phase III is found above 25~T, 
at which the second magnetization step occurs. 
The transition from the paramagnetic phase to the phase III 
takes place at 26~K, which is much higher than the transition temperatures from the paramagnetic to   
phases I and II. At low temperatures, two types of the V sites are observed in phase III in a similar way as 
in phase II, indicating that the high-field phase III exhibits also a heterogeneous spin state. 
Our results show that those Cu spins which are dominantly coupled to the V sites with small 1/$T_2$ 
are responsible for the increase of magnetization at the second step. 
We propose that the second magnetization step is ascribed 
to the ferromagnetic alignment of these Cu spins and discuss possible spin structure in phase III, with particular 
reference to the theories on the anisotropic kagome lattice in the limit $J \gg J^{\prime}$ \cite{Stoudenmire,Schnyder}. 
   
\section{Experiment} 

The $^{51}$V NMR measurements have been performed on a high-quality powder sample 
prepared by the method described in ref.~\citen{HYoshida}. 
The data at high magnetic fields above 18~T were obtained by using 
a 20~MW resistive magnet at LNCMI Grenoble. 
The NMR spectra were recorded at fixed resonance frequency ($\nu _0$) 
by sweeping magnetic field ($B$) in equidistant steps and summing the Fourier transforms 
of recorded spin-echos, which were obtained by using the pulse sequence $\pi /2 - \tau  - \pi /2$ \cite{Clark}. 
They are plotted against the internal field $B_{\mathrm{int}} = \nu _0/\gamma - B$. 
The NMR spectra are labeled by the reference magnetic field $B_0 = \nu _0/\gamma $. 
We determined 1/$T_1$ by fitting the spin-echo intensity $M(t)$ 
as a function of the time $t$ after a comb of several saturating pulses 
to the stretched exponential function $M(t) = M_{eq} - M_0\mathrm{exp}\{-(t/T_1)^{\beta }\}$, 
where $M_{eq}$ is the intensity at the thermal equilibrium. 
This functional form was used to quantify the inhomogeneous distribution of $1/T_1$. 
When the relaxation rate is homogeneous, the value of $\beta $ is close to one.

\section{Results and Discussion} 

\subsection{Temperature dependence of NMR spectra in high magnetic fields} 

\begin{figure}[tb]
\begin{center}
\includegraphics[width=0.8\linewidth]{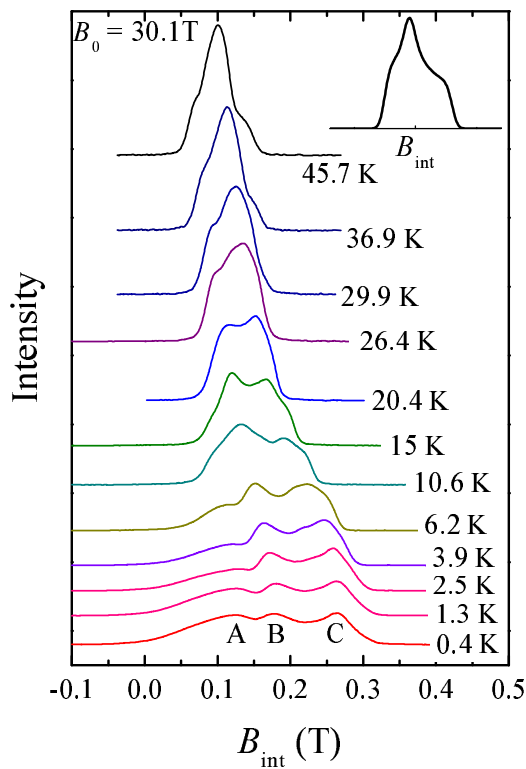}
\end{center}
\caption{(Color online) Temperature dependence of the NMR spectra at $B_0$ = 30.1~T. 
The inset shows an example of calculated powderpattern broadened by a non-axial 
anisotropic magnetic shift.}
\label{fig3}
\end{figure}

Figure~\ref{fig3} shows the temperature dependence of the $^{51}$V NMR 
spectra at $B_0$ = 30.1~T. The spectra above 30~K are compatible with a 
powder pattern due to anisotropic magnetic shifts (inset of Fig.~\ref{fig3}). 
The width due to the quadrupolar interactions is about 0.02~T \cite{MYoshida1},  
much smaller than the observed width in this field range. 
As temperature decreases, the spectral shape begins to change near 26~K. 
Below 20~K, the spectra show two peaks.
Such a structure, which cannot be explained by the anisotropic paramagnetic shift, 
indicates an internal field due to spontaneous antiferromagnetic (AF) moments.
Therefore, our result indicates an AF transition near 26~K at 30~T. 
Similar measurements at 25.6~T indicate the transition near 22~K. 
As shown in Fig.~\ref{fig3}, the spectral width rapidly increases with decreasing 
temperature below 15~K. The spectral shape becomes insensitive to the temperature below 4~K, 
where three broad peaks labeled as $A$, $B$, and $C$ are observed. 

The transition temperature of 26~K is much higher than those between 
the paramagnetic phase and phase I or II below 12 T \cite{MYoshida1}. 
Therefore, this suggests a new magnetic phase at high fields 
which is distinct from phases I and II. 
Indeed, the magnetization data shows the second step near 26~T, 
indicating a field-induced magnetic transition. 
This new high-field phase will be called phase III, as shown in Fig.~\ref{fig2}. 

\begin{figure}[tb]
\begin{center}
\includegraphics[width=0.8\linewidth]{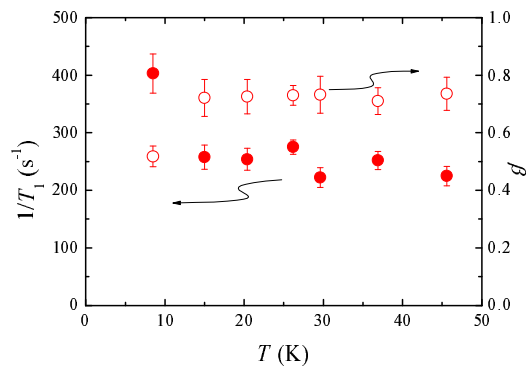}
\end{center}
\caption{(Color online) Temperature dependences of 1/$T_1$ and 
the strech exponent $\beta $ at 30.1~T.}
\label{fig4}
\end{figure}

Figure~\ref{fig4} shows the temperature dependences of 1/$T_1$ and 
the strech exponent $\beta $ at 30.1~T. 
In the temperature range between 15 and 50~K, 1/$T_1$ is independent of temperature. 
In spite of the clear change in the spectral shape, there is no significant anomaly 
in 1/$T_1$ near 26~K. Let us recall that 
while the transition from the paramagnetic phase to phase I is marked by a clear peak 
of 1/$T_1$, only a small anomaly in 1/$T_1$ was observed at 
the transition from the paramagnetic phase to phase II \cite{MYoshida1,MYoshida2}. 
Thus the dynamic signature at the magnetic transition becomes gradually obscured for higher-field phases. 
However, it is still puzzling that, unlike in phase II, 1/$T_1$ no longer decreases below 
the magnetic ordering temperature in phase III.

\begin{figure}[tb]
\begin{center}
\includegraphics[width=0.7\linewidth]{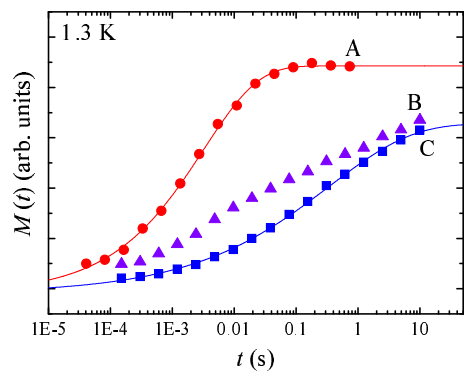}
\end{center}
\caption{(Color online) Recovery curves $M(t)$ measured at positions $A$, $B$, and $C$ 
in Fig.~\ref{fig3} at 1.3~K. 
The solid lines show the fitting to the stretched exponential function 
$M(t) = M_{eq} - M_0\mathrm{exp}\{-(t/T_1)^{\beta }\}$.}
\label{fig5}
\end{figure}

As shown in Fig.~\ref{fig4}, $\beta $ decreases with decreasing temperature 
below 10~K, indicating more inhomogeneous distributions in 1/$T_1$. 
With further decreasing temperature, the distribution of 1/$T_1$ 
becomes too large to determine the representative value of 1/$T_1$. 
In addition, the relaxation rates depend on the spectral position at low temperatures. 
Figure~\ref{fig5} shows the recovery curves $M(t)$ measured at positions $A$, $B$, and $C$ 
in Fig.~\ref{fig3} at 1.3~K. The intensity at $A$ is recovered within 0.1~s, 
while the intensity at $C$ does not saturate even at 10~s. 
These curves can be fit to the stretched exponential function, and 
we obtain 1/$T_1$ = 311 s$^{-1}$ and $\beta $ = 0.49 for $A$ 
and 1/$T_1$ = 2.9 s$^{-1}$ and $\beta $ = 0.33 for $C$. 
The recovery curve at $B$ shows an intermediate behavior between the curves measured at $A$ and $C$. 
Consequently, $M(t)$ at $B$ does not fit to the stretched exponential function. 
These results show that the spectra at low temperatures consist at least of two components. 
This is quite similar to the behavior observed in phase II, 
where the previous NMR measurements show the coexistence of two components characterized by 
the large and small values of 1/$T_2$ \cite{MYoshida2}. 
The two component behavior will be discussed in detail in section 3.3. 
The onset of rapid spectral broadening (Fig.~\ref{fig3}) and the strongly inhomogeneous 
distribution of 1/$T_1$ (Fig.~\ref{fig4}) indicate that there may be another 
transition or crossover at 10 - 15 K. This is shown by the dotted line in Fig.~\ref{fig2}. 

\subsection{Magnetic field dependence of NMR spectra} 

\begin{figure}[tb]
\begin{center}
\includegraphics[width=0.8\linewidth]{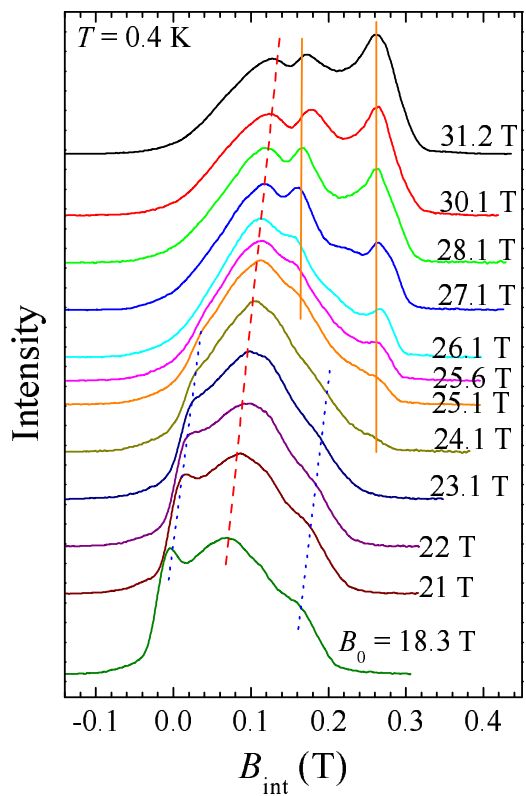}
\end{center}
\caption{(Color online) Magnetic field dependence of the NMR spectra at 0.4~K.
Straight lines are guide to the eye to follow different components (peaks) 
of the spectra. 
}
\label{fig6}
\end{figure}

Figure~\ref{fig6} shows the field dependence of the $^{51}$V NMR spectra at 0.4~K. 
The center of gravity of the spectra shifts to larger values of $B_{\mathrm{int}}$ with 
increasing magnetic field. This shift corresponds to the increase of the magnetization. 
On the other hand, the width and shape of the spectrum are related to the magnetic structure. 
As shown in Fig~\ref{fig6}, the spectral shapes are almost independent of the magnetic field 
in the ranges of 18 - 21~T and above 28~T. However, the spectra above 28~T have a different shape 
from the spectra below 21~T. This indicates that the transition from phase II to phase III 
occurs in the magnetic field region between 21 and 28~T, and the magnetic structure 
in phase III is significantly different from that in phase II. 
In Fig.~\ref{fig6}, we observe that the broad peak indicated by the dashed 
line is present in all the spectra, shifting slightly to higher $B_{\mathrm{int}}$ with increasing magnetic field. 
In phase II, an additional peak and a shoulder indicated by the dotted lines are 
observed at lower and higher $B_{\mathrm{int}}$, respectively. 
They gradually disappears above 22~T. Instead, two additional peaks indicated 
by the solid lines are observed in phase III. 
These two peaks gradually develop above 24~T. As the magnetic field increases from 22 to 28~T, 
there is a gradual transfer of the spectral weight of the peaks 
denoted by the dotted lines to those denoted by the solid lines, 
indicating the coexistence of phases II and III. 

\begin{figure}[tb]
\begin{center}
\includegraphics[width=0.8\linewidth]{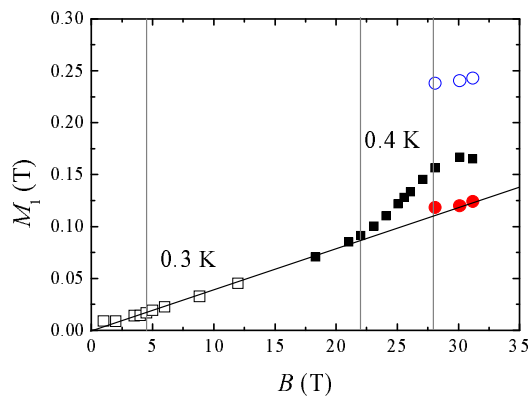}
\end{center}
\caption{(Color online) Magnetic field dependence of the center of gravity of the spectrum $M_1$. 
The experimental temperature is 0.3~K for the data below 15~T and 0.4~K for the data above 15~T. 
The open and solid circles represent $M_1$ for the slow and fast components, respectively 
(see the text).}
\label{fig7}
\end{figure}

In order to investigate the phase boundary in more detail, 
we examine the first and second moments of the spectra defined as, 
\begin{eqnarray}
M_1 &=& \int B_{\mathrm{int}}I(B_{\mathrm{int}})dB_{\mathrm{int}} \nonumber \\
M_2 &=& \int (B_{\mathrm{int}}-M_1)^2I(B_{\mathrm{int}})dB_{\mathrm{int}} 
\end{eqnarray}
where $I(B_{\mathrm{int}})$ is the NMR spectrum normalized as $\int I(B_{\mathrm{int}})dB_{\mathrm{int}}$ = 1. 
Figure~\ref{fig7} shows the magnetic field dependence of $M_1$. 
The data below (above) 15 T were taken at 0.3 K (0.4 K). 
At these sufficiently low temperatures, difference of 0.1 K makes no change 
in the spectral shape. As shown in Fig.~\ref{fig7}, the data below 22~T lie on 
a straight line through the origin. 
Above 22~T, $M_1$ departs from the straight line. The magnetic field dependence 
of $M_1$ in Fig.~\ref{fig7} reproduces the magnetization curve \cite{HYoshida} 
including the second magnetization step at 26 T. 

\begin{figure}[tb]
\begin{center}
\includegraphics[width=0.8\linewidth]{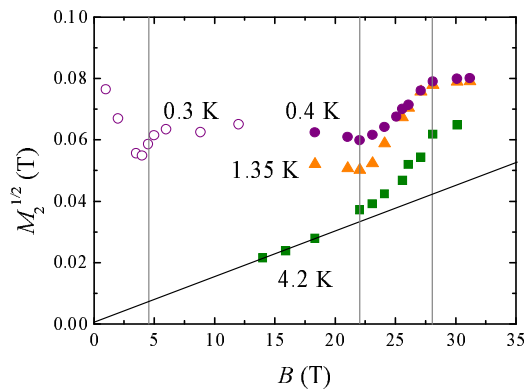}
\end{center}
\caption{(Color online) Magnetic field dependence of square root of the second moment $M_2$. 
The open and solid circles represent the data at 0.3 and 0.4~K, respectively. 
The solid triangles and squares represent the data at 1.35 and 4.2~K, respectively.}
\label{fig8}
\end{figure}

Figure~\ref{fig8} shows the magnetic field dependence of the line width, i.e. square root 
of the second moment $M_2$, at various temperatures. 
At 0.3~K, $\sqrt{M_2}$ decreases with increasing $B$ below 4.5 T (phase I). 
For the range of magnetic field 5 - 22~T and at 0.3 - 0.4~K, $\sqrt{M_2}$ is almost independent of 
$B$, indicating that magnetic structure remains unchanged 
over the entire field range of phase II. 
Above 28~T, $\sqrt{M_2}$ becomes independent of the magnetic field again. 
Thus the region above 28~T should belong to phase III.
For the range of $B$ between 22 and 28~T, $\sqrt{M_2}$ increases monotonically with 
increasing magnetic field. This field region is likely to correspond to the broadened 
transition due to anisotropy in the powder sample. 
The similar broadened transition between phases I and II is also observed around 4.5~T 
with the width of the coexistence region of about 1~T \cite{MYoshida1}. 
In both cases, the width of the transition is about 20 - 25\% of the center field of the transition. 
A likely origin of the broadened transition is the anisotropy of $g$ factor, 
which is estimated to be about 17\% from the electron spin resonance measurements \cite{Ohta}. 

As shown in Fig.~\ref{fig8}, $\sqrt{M_2}$ at 0.4 and 1.35~K show similar behavior above 18 T, 
indicating that the transition field is unchanged at these temperatures. 
On the other hand, $\sqrt{M_2}$ at 4.2~K lie on a straight line through the origin below 18 T. 
This behavior indicates a paramagnetic state, where the internal field is induced by 
the external magnetic field. However, $\sqrt{M_2}$ departs from the straight line above 22~T, 
indicating the appearance of a spontaneous internal field. 
Above 28~T, $\sqrt{M_2}$ becomes independent of the magnetic field. 
Therefore, for the field above 28 T, there must be an AF order even at 4.2~K. 
Thus obtained phase boundaries are plotted by the solid triangles in Fig.~\ref{fig2}. 

\begin{figure}[tb]
\begin{center}
\includegraphics[width=0.8\linewidth]{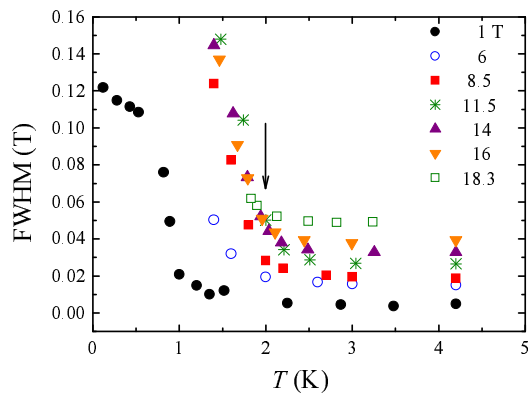}
\end{center}
\caption{(Color online) Temperature dependences of FWHM at various magnetic fields. 
The data below 11.5~T are from in ref. \citen{MYoshida1}.}
\label{fig9}
\end{figure}

In order to establish the phase boundary between phase II and the paramagnetic phase completely, 
the temperature dependence of the full width at half maximum (FWHM) of the spectra 
was measured as shown in Fig.~\ref{fig9} at various magnetic fields. 
Here, the data below 11.5~T are the same as those presented in ref. \citen{MYoshida1}. 
The transition temperatures are determined by the onset of line broadening. 
Above 8.5~T, the onset is insensitive to the magnetic field as 
indicated by the arrow in Fig.~\ref{fig9}. 
The transitions observed by our NMR measurements are summarized in Fig.~\ref{fig2}. 
Phases I and II are characterized by a weak field dependences of the transition temperature. 
Phase III is characterized by much higher transition temperatures 
compared with phases I and II. 

\subsection{Spin structure} 

\begin{figure}[tb]
\begin{center}
\includegraphics[width=0.7\linewidth]{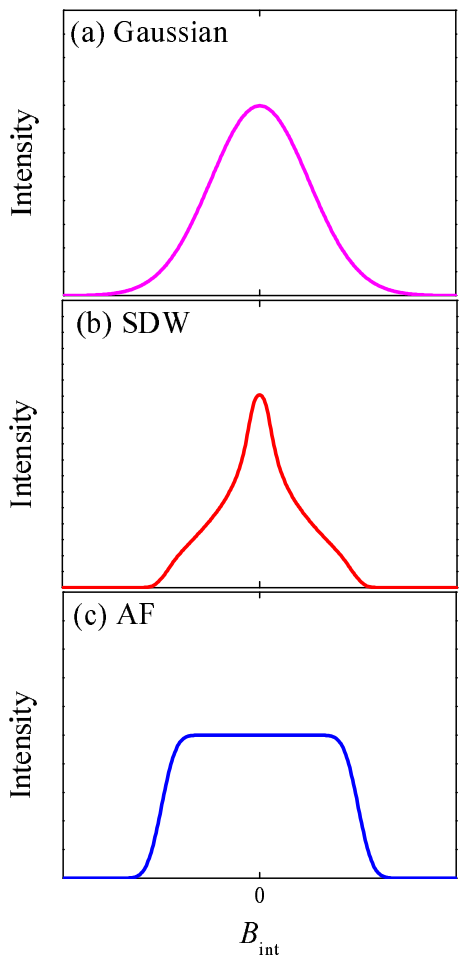}
\end{center}
\caption{(Color online) Typical powder-pattern spectra for three cases, 
(a) Gaussian distribution in the magnitude of the internal field , (b) a SDW state where
the internal field has a sinusoidal modulation, and (c) an N\'{e}el state with a simple two-sublattice 
structure.}
\label{fig10}
\end{figure}

To consider spin structures in the low-temperature phases of volborthite,
we begin with general discussion on the powder-pattern NMR spectra for various magnetic states. 
Figure~\ref{fig10} shows the schematic illustration of the typical powder-pattern for three different antiferromagnetic states. 
The origin of the horizontal axis is set at the center of gravity of the spectrum. 
Fig.~\ref{fig10}(a) shows a Gaussian spectrum which is expected 
if the magnitude of the internal field is randomly distributed. 
Fig.~\ref{fig10}(b) shows a spectrum for a spin density wave (SDW) state \cite{Kontani}, 
where the magnitude of the internal field has a sinusoidal modulation. 
Fig.~\ref{fig10}(c) shows a rectangular spectrum, which is expected 
when the internal field has a fixed magnitude. Examples of this case include a simple N\'{e}el state with a two-sublattice structure and a spiral order with an isotropic hyperfine coupling. For both cases (b) and (c), 
we assume that the direction of the internal field is randomly distributed with respect to the external field, i.e. 
spin-flop transition does not occur. 
As shown in Fig.~\ref{fig10}, both the SDW and Gaussian distribution give a central peak. 
They differ only in the way the intensity vanishes on both sides. 
Since NMR spectra of powder samples are determined by the distribution of the magnitude of 
$B_{\mathrm{int}}$, it may be difficult to distinguish from the NMR spectra 
alone between an SDW order, where $B_{\mathrm{int}}$ 
has a coherent spatial dependence as cos(${\bf Q}\cdot {\bf r}$), and a random distribution 
of $B_{\mathrm{int}}$, in particular if the distribution is not a Gaussian. 
On the other hand, it is easy to distinguish those ordered states with spatial modulation 
in the magnitude of moments such as SDW from the ones with a fixed magnitude of moments. 
The spectra observed in phase I belongs to the former case as reported in ref.~\citen{MYoshida1},  
i.e. the magnitude of the internal field is spatially modulated. 

\begin{figure}[tb]
\begin{center}
\includegraphics[width=0.8\linewidth]{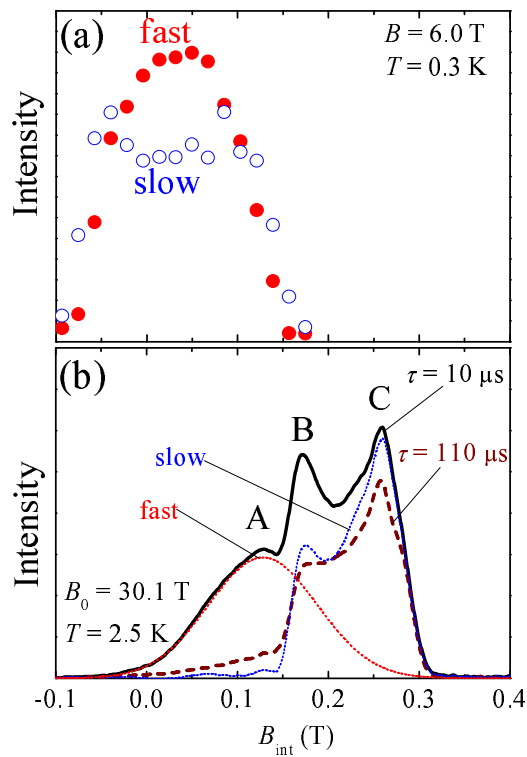}
\end{center}
\caption{(Color online) (a) NMR spectra for the fast (solid circles) and slow (open circles) 
components at 0.3~K in the magnetic field of 6.0~T. 
(b) NMR spectra at $\tau $ = 10 (thick solid line) and 110~$\mu $s (thick dashed line) 
in the magnetic field of 30.1~T. The red (blue) dotted line represents the NMR spectra 
for the fast (slow) component.}
\label{fig11}
\end{figure}

In phase II, the previous NMR results show coexistence of two types of V sites, 
V$_f$ and V$_s$ sites, which are characterized 
by the large (fast) and small (slow) relaxation rates, respectively \cite{MYoshida2}. 
The spin-echo decay curve in phase II was well fit to the two-component 
function $I(\tau ) = A_f\mathrm{exp}(-2\tau /T_{2f}) + A_s\mathrm{exp}(-2\tau /T_{2s})$. 
The coefficients $A_f$ and $A_s$ are proportional to the number of V$_f$ and V$_s$ sites 
within the frequency window covered by the exciting rf-pulse. 
Figure~\ref{fig11}(a) shows the spectra for the V$_f$ and V$_s$ sites at 6 T and 0.3 K 
reported in ref. \citen{MYoshida2}. 
These spectra are obtained by plotting $A_f$ and $A_s$ at various frequencies $\nu_0 $ 
with the fixed $B$ = 6.0~T, after the two-component fitting has been done. 
The spectrum of the V$_f$ sites resembles a Gaussian. 
Therefore, the V$_f$ sites in phase II have a similar feature to the V sites in phase I. 
These V sites should be surrounded by Cu moments with spatially varying magnitude. 
On the other hand, the spectrum of the V$_s$ sites shows a rectangular-like shape. 
indicating that the internal field at the V$_s$ sites is more homogeneous, 
compared with that at the V$_f$ sites. The spin structure of the V$_s$ sites can be 
a two-sublattice N\'{e}el order or a spiral order with an isotropic hyperfine coupling. 
When the hyperfine coupling tensor is slightly anisotropic, a spiral order leads 
to a small modulation of the internal field, resulting in a trapezoidal spectrum. 
In reality, this may be difficult to distinguish from a rectangular one. 
Thus we cannot select a unique spin structure. 

The multi-component behavior is observed also in phase III by the 1/$T_1$ measurements at low temperatures  
as shown in Fig. \ref{fig5}. These components also exhibit different 
spin-echo decay rates. In Fig.~\ref{fig11}(b), two spectra at 30.1~T and 2.5~K 
are displayed, one for $\tau $ = 10 $\mu $s and the other for $\tau $ = 110 $\mu $s. 
The spectrum for $\tau $ = 10 $\mu $s is the same as the one shown in Fig.~\ref{fig3}, 
and the peaks are labeled as $A$, $B$, and $C$. The intensity of the peak $A$ markedly 
decreases with increasing $\tau $ from 10 to 110 $\mu $s. 
That is, the $A$ component has large 1/$T_2$ as well as large 1/$T_1$. 
The left half of the peak $A$ of the spectrum for $\tau $ = 10 $\mu $s can be fit to a 
Gaussian labeled as ``fast'', which disappears in the spectrum for $\tau $ = 110 $\mu $s. 
By subtracting the Gaussian component from the spectrum for $\tau $ = 10 $\mu $s, 
we obtain a rectangular component shown by the dotted line 
labeled as ``slow''. Then, the ``slow'' component 
is quite similar to the spectrum for $\tau $ = 110 $\mu $s, 
indicating that the two-component description is valid also in phase III. 

The common feature of phases II and III are two types of V sites, 
the V$_f$ and V$_s$ sites, of similar characteristics. 
At the V$_f$ sites, the spectral shapes resembles a Gaussian and the relaxation 
rates are large. On the other hand, the V$_s$ sites have a rectangular-like spectral shape 
with relatively small relaxation rates. However, the average values of $B_{\mathrm{int}}$ 
for the two V sites are quite different for phases II and III as shown in Figs.~\ref{fig11}(a) and (b). 
In going from phase II to phase III, the center of gravity of the whole spectrum 
shifts to higher values of $B_{\mathrm{int}}$ due to increase of the magnetization. 
What is more interesting is that the relative positions of the two components  
change substantially between phases II and III.
In phase II, the centers of gravity of the fast and slow components are located 
almost at the same position, 
as shown in Fig.~\ref{fig11}(a). This shows that the two types of Cu spins, 
each of which is dominantly coupled to either V$_f$ or V$_s$ sites, have almost 
the same field-induced averaged magnetization. 
On the other hand, in phase III, the center of gravity of the slow component is 
located at a much larger value of $B_{\mathrm{int}}$ than that of the fast component. 
Therefore, the averaged magnetization coupled to 
the V$_s$ sites is much larger than that coupled to the V$_f$ sites in phase III. 
The centers of gravity of the fast and slow components above 28~T are shown in Fig.~\ref{fig7}. 
The former is on the straight line extrapolated from phase II, 
while the latter is located at much larger $B_{\mathrm{int}}$. 
Therefore, we conclude that the Cu spins coupled to the V$_s$ sites are 
responsible for the increase of magnetization at the second step. 

The center of gravity $M_1$ is related to the magnetization $M$ by the relation 
$M = M_1/A_{\mathrm{hf}}$, where $A_{\mathrm{hf}}$ is the coupling constant 
determined in the paramagnetic phase $A_{\mathrm{hf}}$ = 0.77~T/$\mu _B$ \cite{Bert,MYoshida1}. 
Indeed, the center of gravity of the whole spectrum at 31~T, 
$M_1$ = 0.165~T, gives the magnetization of 0.21~$\mu _B$ in good agreement with the 
measured magnetization of 0.22~$\mu _B$ \cite{HYoshida}. 
By using this relation, the averaged magnetization coupled to the V$_f$ (V$_s$) sites, 
$M_{\mathrm{fast}}$ ($M_{\mathrm{slow}}$), at 31 T are estimated to be 0.16 (0.32~$\mu _B$). 
The integrated intensities of the spectra of the fast and slow components are almost the same. 
This means that numbers of the two sites $N_f$ and $N_s$ are nearly equal. 
More precise determination of the ratio $N_f:N_s$ would require precise knowledge of the 
frequency dependence of 1/$T_1$ and 1/$T_2$. 

In phase II, a heterogeneous spin state was proposed \cite{MYoshida2}, 
in which the V$_f$ and V$_s$ sites spatially alternate, because of 
magnetic superstructure on Cu sites. 
Arguments were given in ref.~\citen{MYoshida2} why this is more likely 
than the macroscopic phase separation. 
First, the ratio $N_f:N_s$ should vary as a function of $B$ and $T$, 
if the V$_f$ and V$_s$ sites belong to distinct phases. 
Experimentally, however, $N_f$ and $N_s$ are nearly equal irrespective of $B$ and $T$. 
Second, the V$_f$ and V$_s$ sites show similar $T$-dependence of 1/$T_2$ \cite{MYoshida2}. 
This means that the two types of Cu spins coupled to the V$_f$ and V$_s$ sites, 
respectively, are interacting in a certain way to establish a common $T$-dependence. 
Again, this is not possible if they belong to different phases. 

\begin{figure}[tb]
\begin{center}
\includegraphics[width=0.8\linewidth]{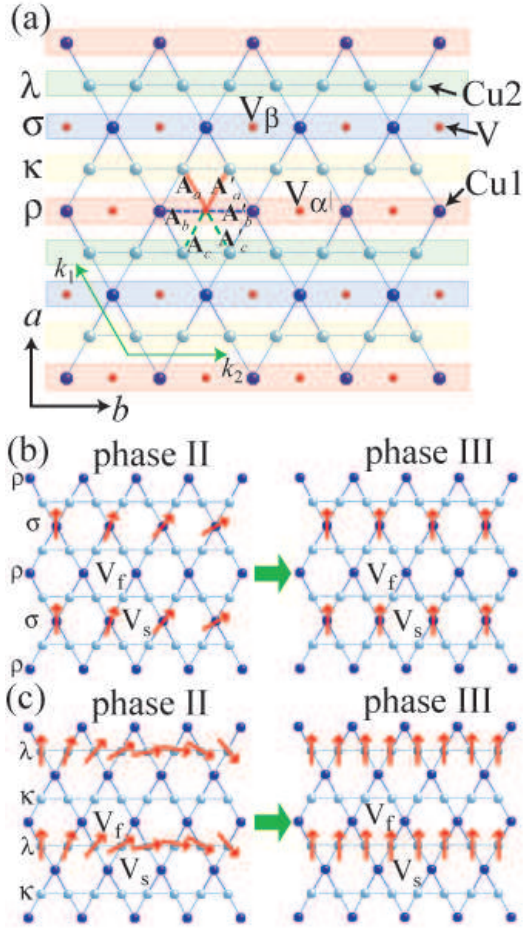}
\end{center}
\caption{(Color online) Possible magnetic structures in volborthite projected onto the $a$-$b$ plane.
(a) Superstructure along the ${\bf k}_1$ direction provides 
two inequivalent V sites, V$_{\alpha }$ and V$_{\beta }$, and four inequivalent Cu sites, 
Cu1$_{\rho}$, Cu1$_{\sigma}$, Cu2$_{\kappa}$, and Cu2$_{\lambda }$. 
(b) If the coupling tensors ${\bf A}_a$, ${\bf A}_b$, and ${\bf A}_c$ are largely isotropic 
and have similar magnitudes, different relaxation behavior at the V$_{f}$ and V$_{s}$ sites should be ascribed
to the distinct spin states at the Cu1$_{\rho}$ and Cu1$_{\sigma}$ sites.    
(c) If ${\bf A}_a$ is the dominant coupling, different fluctuations at the Cu2$_{\kappa}$
and Cu2$_{\lambda }$ sites lead to distinction between V$_{f}$ and V$_{s}$. 
}
\label{fig12}
\end{figure}

Although it is difficult to determine the superstructure from 
the NMR data on a powder sample alone, 
plausible structures were discussed in the previous paper as follows \cite{MYoshida2}. 
The most simple way to generate two types of V sites with the same abundance is to form a 
superstructure doubling the size of the unit cell. 
On the ideal kagome lattice, the three directions 
${\bf k}_1$, ${\bf k}_2$, and ${\bf k}_1$ + ${\bf k}_2$ (Fig.~\ref{fig12} a) 
are all equivalent. However, in volborthite, the structure is uniform along ${\bf k}_2$ 
and inequivalent Cu1 and Cu2 sites alternate along ${\bf k}_1$. 
Then it is more likely that the superstructure develops 
along ${\bf k}_1$. Consequently, each of the Cu1 and Cu2 sites are divided 
into two inequivalent sites Cu1$_{\rho}$, Cu1$_{\sigma}$ and Cu2$_{\kappa}$, Cu2$_{\lambda }$, 
as shown in Fig.~\ref{fig12}(a). It follows that there are two types of the V sites, 
V$_{\alpha }$ and V$_{\beta }$, as shown in Fig.~\ref{fig12}(a). 
They should correspond to the V$_f$ and V$_s$ sites. It should be emphasized that
distinct spin states on the crystallographically inequivalent Cu1 and Cu2 sites 
do not account for our observation of two different V sites that are otherwise
crystallographically equivalent. Our results can be accounted for only by a symmetry breaking superstructure. 

Since a V site is located approximately above or below the center 
of Cu hexagon of the kagome lattice, the dominant source of the 
hyperfine field at V nuclei should be confined within the six Cu spins on a hexagon. 
Because of the distorted structure of volborthite, 
there are three distinct hyperfine coupling tensors 
${\bf A}_a$, ${\bf A}_b$, and ${\bf A}_c$ as shown in Fig.~\ref{fig12}(a). 
The coupling tensors to the other three spins 
${\bf A}_a'$, ${\bf A}_b'$, and ${\bf A}_c'$ are generated by the mirror 
reflection perpendicular to the $b$ axis at the V site. 
${\bf A}$ and ${\bf A}'$ are the same hyperfine tensors but oriented differently, 
and they will be effectively different only if the tensor is anisotropic. 
The coupling constant $A_{\mathrm{hf}}$ determined in the paramagnetic state 
should be equal to the average of the diagonal components of the sum of these six tensors. 
The hyperfine fields at the V$_{\alpha }$ and V$_{\beta }$ sites are written as 
${\bf B}_{\alpha } = ({\bf A}_a{\bf s}_{\kappa } + {\bf A}_a'{\bf s}_{\kappa }') 
+ ({\bf A}_b{\bf s}_{\rho } + {\bf A}_b'{\bf s}_{\rho }') 
+ ({\bf A}_c{\bf s}_{\lambda } + {\bf A}_c'{\bf s}_{\lambda }') $ 
and ${\bf B}_{\beta } = ({\bf A}_a{\bf s}_{\lambda } + {\bf A}_a'{\bf s}_{\lambda }') 
+ ({\bf A}_b{\bf s}_{\sigma } + {\bf A}_b'{\bf s}_{\sigma }') 
+ ({\bf A}_c{\bf s}_{\kappa } + {\bf A}_c'{\bf s}_{\kappa }') $, 
respectively, where ${\bf s}_{\epsilon }$ and ${\bf s}_{\epsilon }'$ 
denote two neighboring spins on the same type of sites 
(i.e., $\epsilon$ stands for Cu1$_{\rho}$, Cu1$_{\sigma}$, Cu2$_{\kappa}$, or Cu2$_{\lambda }$). 

Both 1/$T_1$ and 1/$T_2$ are determined by the time correlation 
function of the hyperfine field \cite{Slichter}. 
Since the hyperfine coupling to the uniform magnetization $A_{\mathrm{hf}}$
is largely isotropic \cite{MYoshida1}, we expect this is also the case 
for the individual coupling tensors ${\bf A}_a$, ${\bf A}_b$, and ${\bf A}_c$. 
If they have similar magnitudes, the coupling to the Cu2 sites is identical for the 
two V sites, and the different relaxation rates must be ascribed to the difference 
in the time correlation functions of the Cu1 spins ${\bf s}_{\rho }$ and ${\bf s}_{\sigma }$. 
The model proposed in ref.~\citen{MYoshida2} is shown in Fig.~\ref{fig12}(b). 
The Cu1$_{\sigma }$ sites develop a long range magnetic order 
with a uniform magnitude of moments. The spin fluctuations at these sites should have 
a small amplitude, leading to the slow relaxation at the V$_s$ sites. 
On the other hand, the Cu1$_{\rho}$ sites show a modulation in the magnitude 
of the ordered moments. The absence of fully developed static moments would 
allow unusually slow fluctuations responsible for the large 1/$T_2$ at the V$_f$ sites. 

In phase II, the spectra of the V$_f$ and V$_s$ sites have almost the same $M_1$, which is proportional 
to the field. Therefore, all Cu sites have both an AF component and a field-induced uniform component. 
In phase III, $M_1$ is largely shifted only at the V$_s$ sites. This shift can be explained by ferromagnetic
alignment of the Cu1$_{\sigma}$ sites as shown in the right panel of Fig.~\ref{fig12}(b). As shown in
Fig.~\ref{fig11}, the width of the spectrum of the V$_s$ sites in phase III is narrower than that in phase II. 
This decrease of the width is also understood by the ferromagnetic saturation of the Cu1$_{\sigma }$ sites. 
Because the ferromagnetic moment is always parallel to the magnetic field, the spectral width is determined by
the anisotropy of the hyperfine coupling. Indeed, the spectral shape of the V$_s$ sites in phase III is similar
to a paramagnetic powder pattern. In contrast, the width in phase II should be dominantly determined by the AF
component. On the other hand, the position and the width of the spectrum of the V$_f$ sites do not change
substantially at the transition between phases II and III. This indicates that the 
Cu1$_{\rho}$ spins keep spatial modulation of the magnitude of moment in phase III. We should stress that the width
of the V$_s$ spectrum is much smaller than the shift $M_1$ in phase III, consistent with the largely 
isotropic hyperfine coupling in the paramagnetic phase \cite{MYoshida1}. 

Our results show interesting correspondence with the theory on the anisotropic 
kagome model in the limit $J \gg J^{\prime}$ \cite{Schnyder,Stoudenmire}. 
Schnyder, Starykh, and Balents studied the ground state of this model in zero magnetic 
field. They found that the Cu1 sites develop either a ferromagnetic order or a spiral order with 
a long wave length with the full moment, while the Cu2 sites have only a small ordered 
moment \cite{Schnyder}. The spin fluctuations at the Cu2 sites associated with the large energy 
scale $\sim J$ should also lead to a small contributions to 1/$T_1$ and 1/$T_2$ at the V sites.  
The irrelevance of the Cu2 sites for both the static internal field and the dynamic 
relaxation rates at the V sites is consistent with our model that the difference   
between the V$_f$ and V$_s$ sites are ascribed to the two types of the Cu1 sites. 
It should be noted, however, that the theory does not account for the superstructure 
which produces the different V$_f$ and V$_s$ sites.  

The similarity between our model shown in Fig.~\ref{fig12}(b) and 
the theoretical prediction on the anisotropic kagome model becomes more apparent in high magnetic fields. 
A moderate magnetic field should easily drive the small-$Q$ spiral order 
of the Cu1 sites into a ferromagnetic state. 
This situation is consistent with our observation of 
induced ferromagnetism of Cu1$_{\sigma}$ spins in phase III. However, our results 
indicate that only a half of the Cu1 sites undergoes 
ferromagnetic saturation across the second magnetization step at 26 T. 
It appears most likely that the other half of the Cu1 sites (the Cu1$_{\rho}$ sites) 
get fully polarized across the third step at 46 T. Stoudenmire and Balents investigated the spin structure of 
the Cu2 sites for the same anisotropic kagome model in high magnetic fields, where 
Cu1 moments are fully aligned by the field \cite{Stoudenmire}. They found that the Cu2 sites 
show a quantum phase transition from a ferrimagnetic order along the field direction to an 
antiferromagnetic order perpendicular to the field with increasing field. A remarkable aspect of the 
theory is the prediction for a magnetization plateau at 2/5 of the saturation magnetization in 
the ferrimagnetic region immediately below the transition field. 

The 2/5 plateau has been actually observed in the recent magnetization measurements above 60 T \cite{Okamoto2}.
Therefore, we further examine if such a scenario is compatible with our experimental
results. When the Cu1 sites exhibit the full moment, $\langle s_z \rangle$ = 1/2, 
in the plateau region, the Cu2 sites should have the averaged moment $\langle s_z \rangle$ = 1/20 
in order to account for the total magnetization of 2/5 of the saturation. 
If we assume that all the hyperfine coupling tensors 
${\bf A}_a$, ${\bf A}_b$, and ${\bf A}_c$ are isotropic and have the same magnitude, 
$A_a = A_b = A_c = A_{\mathrm{hf}}$/6 = 0.13 T/$\mu_B$, the contribution from the four Cu2 sites to $B_{\mathrm{int}}$ is 0.055 T at all V sites 
in the plateau region. Here, we used the averaged $g$ value of 2.15.\cite{Ohta,Zhang} 
This contribution to $B_{\mathrm{int}}$ will be reduced to 0.028 T at 30 T, assuming that 
the magnetization at the Cu2 sites is proportional to $B$. In addition, 
the full moments at the Cu1$_{\sigma}$ sites produce $B_{\mathrm{int}}$ = 0.28 T at the V$_s$ sites. 
The sum of these gives $B_{\mathrm{int}}$ = 0.30 T, which is 
reasonably close to the observed $M_1$ (= 0.24 T) at the V$_s$ sites at 30 T (Fig. ~\ref{fig7}). 
On the other hand, $M_1$ = 0.12 T at the V$_f$ sites at 30 T 
corresponds to $\langle s_z \rangle$ = 0.17 at the Cu1$_{\rho }$ sites. 
The summation of the magnetizations at the Cu2, Cu1$_{\sigma}$, and Cu1$_{\rho}$ sites 
amounts to the total magnetization of  0.28 $\mu_B$ at 30 T, 
which is in reasonable agreement with the experimental magnetization of 0.22 $\mu_B$. 
Therefore, this model is semiquantitatively compatible with the NMR and magnetization measurements. 

Let us remark that the V$_s$ and V$_f$ sites show extremely different relaxation rates  
in phase III. As shown in Fig.~\ref{fig5}, the V$_s$ sites (line $C$) show two orders of magnitude 
smaller 1/$T_1$ than the V$_f$ sites (line $A$). The spectra in Fig.~\ref{fig11}(b) for different 
values of $\tau$ indicate orders of magnitude difference in 1/$T_2$. In contrast, 
the difference in 1/$T_2$ between the V$_f$ and V$_s$ sites is only a factor of four  
in phase II \cite{MYoshida2}. In fact the extremely small relaxation rates at the V$_s$ sites in phase III 
is what should be expected if the moments at the Cu1$_{\sigma}$ sites are completely saturated by magnetic fields 
and the contribution from the Cu2 sites is heavily suppressed for $J \gg J^{\prime}$. 
On the other hand, in phase II, the spins at the Cu1$_{\sigma}$ sites have finite fluctuations even though 
they show a conventional magnetic order. The anomalous fluctuations of the Cu1$_{\rho}$ spins can then 
couple to the fluctuations of the Cu1$_{\sigma}$ spins, leading to the similar temperature dependence of 
1/$T_2$ at the V$_f$ and V$_s$ sites \cite{MYoshida2}.

The above discussion is based on the assumption that all 
the hyperfine coupling tensors are largely isotropic and have 
similar magnitudes. However, the low symmetry of the volborthite structure may result 
in large difference in the hyperfine coupling. In particular, ${\bf A}_a$ and ${\bf A}_c$ 
involve significantly different V-O-Cu hybridization paths. 
If one of them, say ${\bf A}_a$, is the dominant coupling, 
the different relaxation rates for the two V sites must be 
ascribed to the different dynamics of ${\bf s}_{\lambda }$ and 
${\bf s}_{\kappa }$. An example of this case is shown in Fig.~\ref{fig12}(c). 
However, there has been no microscopic theory which supports such a scenario. 

\section{Summary} 
We have presented the $^{51}$V NMR results of the $S$ = 1/2 distorted kagome lattice volborthite. 
Following the previous experiments in magnetic fields below 12 T, 
the NMR measurements have been extended to higher fields up to 31~T. 
In addition to the already known two ordered phases (phases I and II), 
we found a new high-field phase above 25~T, at which the second 
magnetization step occurs. 
This high-field phase (phase III) has the transition temperature of about 26~K, 
which is much higher than those of phase I (1 K) and phase II (1 - 2 K). 
At low temperatures, two types of the V sites are observed with different relaxation 
rates and line shapes in phases II and III. Our results indicate that both phases 
exhibit a heterogeneous spin state, where an ordered state with non-uniform magnitude of 
moment with anomalous fluctuations alternates with a more conventional order with a fixed magnitude
of the ordered moment. The latter group of spins are responsible for the increase of magnetization 
at the second step between the phase II and phase III. 
We proposed a possible spin structure in phase III compatible with the NMR and magnetization measurements. 
Our model has common features with the theories on the anisotropic kagome lattice 
in the limit $J \gg J^{\prime}$ \cite{Schnyder,Stoudenmire}, 
although the formation of heterogeneous superstructure was not predicted by the theories. 

\section*{Acknowledgment}

We thank F. Mila for stimulating discussions. 
This work was supported by MEXT KAKENHI on Priority Areas 
``Novel State of Matter Induced by Frustration'' (No. 22014004), 
JSPS KAKENHI (B) (No. 21340093), the MEXT-GCOE program, 
and EuroMagNET under the EU contract NO. 228043.

\end{document}